\documentclass[a4paper,10pt,twoside]{cpc-hepnp}

\usepackage{multicol}
\usepackage{graphicx}
\usepackage{booktabs}
\usepackage{amssymb,bm,mathrsfs,bbm,amscd}
\usepackage[tbtags]{amsmath}
\usepackage{lastpage}
\usepackage{CJK}

\begin{document}
\begin{CJK*}{GBK}{song}

\fancyhead[c]{\small To be published in Chinese Physics C (2020)}
\fancyfoot[C]{\small \thepage}


\title{Investigation of neutron density distribution of $^{208}$Pb nucleus when the proton density is constrained to its experimental distribution
}

\author{%
     A. R. Abdulghany\email{abdulghany@sci.cu.edu.eg}%
}
\maketitle

\address{%
  Physics Department, Faculty of Science, Cairo University, Giza, Egypt
}

\begin{abstract}
In this study, two novel improvements for the theoretical calculation of the neutron distributions are presented. First, the available experimental proton distributions are used as a constraint rather than inferred from the calculation. Second, the recently proposed distribution formula, d3pF, is used for the neutron density, which is more detailed than the usual shapes, for the first time in nuclear structure calculation. A semi-microscopic approach for binding energy calculation is considered in this study, however, the proposed improvements can be introduced to any other approach. The ground state binding energy and neutron density distribution of $^{208}$Pb nucleus are calculated by optimizing the binding energy considering three different distribution formulae. The implementation of the proposed improvements leads to a qualitative and quantitative improvement in the calculation of the binding energy and neutron density distribution. The calculated binding energy agrees with the experimental value, and the calculated neutron density shows fluctuations within the nuclear interior, which agrees with the predictions of self-consistent approaches.
\end{abstract}




\begin{multicols}{2}

\section{Introduction}
Since Rutherford discovered the atomic nucleus \cite{ref1} there is sustained interest in research on the proton and neutron density distributions in finite nuclei. An accurate knowledge of density distributions is crucial for understanding the fundamental properties of nuclear matter and the nature of nuclear force. The proton differs from the neutron in nature and interactions, and this difference is reflected in the distributions of protons and neutrons within the nucleus. In general, the distribution of protons differs from the distribution of neutrons, qualitatively and quantitatively, especially in the range of very heavy and superheavy nuclei, where increasing the neutrons/protons ratio is necessary to maintain the stability of the nucleus. The difference between neutron and proton distributions is always abstracted as the difference between the neutron and proton distributions root-mean-square (rms) radii, known as the ``neutron skin thickness''. The neutron skin thickness is fundamental for some crucial problems in modern nuclear physics and astrophysics. It has been found that including the neutron skin thickness in the $\alpha$-decay calculation improves the calculation of $\alpha$-daughter potential and the prediction of $\alpha$-decay half-lives and $\alpha$ preformation probability \cite{rev+1,rev+2}. Similar effects have been reported in the calculation of the $\alpha$-decay of neutron-deficient nuclei that have ``proton skin'' instead of the neutron skin \cite{rev+3}. Not only is the value of neutron skin thickness important in the study of decay, but it has also been reported that the changes in neutron  skin thickness from parent to daughter nuclei consistently correlate with the observed half-lives \cite{rev+4}. An accurate estimation of neutron and proton distributions is also crucial in the study of asymmetric nuclear matter, which is a contact area between nuclear physics and astrophysics \cite{rev+5,rev+6,rev+7}.

In the 1950's, Hofstadter pioneered electron scattering experiments \cite{ref2}, which provided, in addition to subsequent refinements, an accurate description of electric charge distributions of stable isotopes \cite{ref3,ref4,ref5,ref6}, and short-lived isotopes \cite{ref7,ref8}. The most accurate form of charge densities is provided as a model independent distributions in terms of Fourier-Bessel coefficients \cite{ref3,ref4,ref5}. Several analytical formulae with a diffused edge may be used to represent the charge densities involving fewer parameters than the number of Fourier-Bessel coefficients. Such forms are commonly used in both nuclear reaction and nuclear structure studies.
For most applications, two-parameter Fermi (2pF) distributions and three-parameter Fermi (3pF) distributions are acceptable approximations for nuclear charge and matter distributions \cite{ref9,ref10,ref11}. The 2pF and 3pF distributions are given by equations \ref{eq1} and \ref{eq2}, respectively, where a is the diffuseness parameter, $R$ is the radius of the nucleus and $w$ is the central depression parameter. The central density $\rho_0$ is determined by the normalization to the number of protons (Z) or neutrons (N).

\begin{equation}
\label{eq1}
\rho(r)=\frac{\rho_0}{1+e^{(r-R)/a}}.
\end{equation}

\begin{equation}
\label{eq2}
\rho(r)=\frac{\rho_0 \left[1+w(r^2/R^2) \right] }{1+e^{(r-R)/a}}.
\end{equation}

Although the 2pF distribution is the most widely used in the study of nuclear structure, reactions, and decay, using 3pF distribution improves the calculation of binding energy \cite{ref12} and alpha decay half-life time and preformation probability \cite{ref13}. Recent studies show that fitting of the experimental charge distributions to 3pF distribution does not provide a significant improvement over the fitting to 2pF \cite{ref10,ref14}. This is because both functions were unable to describe fluctuation in density within the nuclear interior that appears in the experimental distributions. The recently proposed double 3pF distribution (d3pF), which allows for density fluctuation, fits the experimental charge densities with significant improvement in accuracy over other commonly used formulae \cite{ref14}. The d3pF distribution, given by equation \ref{eq3}, is composed of two 3pF terms, one with larger radius parameter, in order to describe the tail of the density, and the second has smaller radius, to describe density fluctuation at the nucleus interior.

\begin{equation}
 \label{eq3}
\rho(r)= \rho_0 \sum\limits_{i=1,2}  \frac{\delta_i\left[1+w_i(r^2/R_i^2) \right] }{1+e^{(r-R_i)/a_i}},
\\
\delta_1+\delta_2 = 1,
\end{equation}
where $\delta_i$'s are the weights of the two 3pF parts. This function has seven independent parameters since the density distribution should verify the normalization condition.

Although the charge distributions were measured accurately, the data for neutron distributions are still deficient. The study of neutron distributions have attracted interest of researchers because of its fundamental importance, it determines the nuclear drip lines and stability regions \cite{ref15}, gives rise to special structures and phenomena in some isotopes \cite{ref16,ref17,ref18}, and is germane to the structure of neutron stars \cite{rev+7,ref19,ref20}. There have been many efforts worldwide to develop and implement experiments to characterize the neutron distributions in nuclei using different techniques. The neutron distributions have been probed mostly by hadron scattering \cite{ref22} , $\alpha$-scattering \cite{ref23}, nuclear pion photoproduction \cite{ref24}, or electroweak electron scattering \cite{ref25}. Because of complexity of the strong force, the hadronic probes require model assumptions to deal with it, and there results are model-dependent. Electroweak probe was introduced as a model-independent probe of neutron distributions \cite{ref26,ref27}, it mainly characterizes the distribution of weak charge in nuclei, which is mainly of the neutrons since the weak charge of the proton is about 7\% its value for the neutron \cite{ref28}. The experimental efforts in the study of the neutron distribution are directed extensively to measure the distribution rms radii \cite{ref26,ref29,ref30} and the neutron skin thickness \cite{ref26,ref31}. The measured neutron distribution rms radius of $^{208}$Pb ranges from 5.6 fm to 5.94 fm \cite{rev1,rev2} and neutron skin thickness ranges from 0.09 to 0.49 fm \cite{ref24,rev1,rev2}. This uncertainty may originate from the limitations of the measurements in addition to statistical and systematic errors. In an attempt to explore the density distribution, the recent coherent pion-photoproduction experiment provided the neutron density of $^{208}$Pb by fitting pion-nucleus scattering data to the 2pF distribution with $R=6.7 \pm 0.03$ fm and $a= 0.55 \pm 0.03$ fm \cite{ref24}.

As the experimental studies have achieved success in the characterization of proton density in nuclei, many theoretical approaches have been developed to provide information about nuclear structure such as binding energy, deformations, proton and neutron density distributions and exotic nuclear structures. There are different levels of theoretical calculations, starting from the pure fundamental ab initio methods, to the pure phenomenological methods such as liquid drop model. In between the two extremes of the calculation levels, there are some approaches with simplified potentials are proposed to deal the many-particle problem, e.g., self-consistent mean-field models \cite{ref32} or shell model. The most notable alternative to the self-consistent method is the semi-microscopic methods \cite{ref33,ref34} with strutinsky shell-correction \cite{ref35,ref36} in which the energy of a nucleus is considered to be the sum of macroscopic and microscopic parts. In addition to its simplicity, the semi-microscopic technique shows vast success in studying the nuclear structure \cite{ref12,ref13,ref37,ref38,ref39,ref40} and $\alpha$-decay \cite{ref41}. The semi-microscopic approach used in this study have used successfully to reproduce data of the latest discovered superheavy element $^{294}$Og and other Og isotopes with accuracy in line with experimental data \cite{ref38}. It also predicts the masses and deformations for heavy and superheavy nuclei with accuracy in line with the prevalent microscopic and semi-microscopic models \cite{ref12,ref39,ref40}.

\section{Theoretical Framework}
In this study, the total energy is evaluated in the framework of the semi-microscopic approach prescribed in \cite{ref12,ref39}. In such method, the macroscopic part of the total energy is considered by the energy density functional, with the Skyrme force SkM*, as a function of the nucleon densities $\rho_i$ and the kinetic energy densities $\tau_i$ \cite{ref42} in the form,
\begin{equation}
\label{eq4}
\mathcal{H}(\textbf{r}) = \frac{\hbar^2}{2m}[\tau_p (\textbf{r})+\tau_n (\textbf{r})]+\mathcal{H}_{\text{\text{Sk}}}(\textbf{r})+\mathcal{H}_{\text{Coul}}( \textbf{r}),
\end{equation}
where $\mathcal{H}_{Sk}(r)$ is the nuclear energy density and $\mathcal{H}_{Coul}(r)$ is the Coulomb energy density. The macroscopic part of the total energy ($E$) is given by the volume integral of the total energy density.
\begin{equation}
\nonumber
E =\int \mathcal{H} \-\bold{dr}.
\end{equation}

The microscopic contribution is considered with the Strutinsky's shell and pairing correction method. The shell-correction energy is obtained as the difference between the sum of occupied energy levels, obtained from the Woods-Saxon single particle Hamiltonian, and the corresponding sum obtained by Strutinsky's smoothing procedure \cite{ref35,ref43}. The Barden Cooper-Scheffer (BCS) approach is used to calculate residual pairing correction energy following the procedure of \cite{ref44}.

In order to consider the finer details of the experimental proton density, we use the model independent Fourier-Bessel expansion \cite{ref3,ref4,ref5} given by,
\begin{equation}
\label{eq5}
\rho(r) =
\left\{
	\begin{array}{ll}
		\sum_{v}a_vj_0(v\pi r/R)  &  ,  r\leq R \\
		0                         &  ,  r > R
	\end{array}
\right.
\end{equation}
where $j_0$ is the zero order spherical Bessel function, $a_v$'s and $R$ are Fourier-Bessel coefficients deduced from experiments and given in data compilation \cite{ref4}. The 2pF, 3pF and d3pF distributions, as given by equations \ref{eq1}, \ref{eq2} and \ref{eq3}, respectively, are used to parameterize the neutron density distribution in the calculation of total energy.

\section{Results and Discussion}
In the present study, the binding energy of $^{208}$Pb is calculated assuming that the proton density distribution is constrained to its experimental distribution. The neutron density distribution is considered in the form of parameterized analytical distributions. The semi-microscopic model based on energy density functional with the Skyrme force and the Strutinsky's shell and pairing correction is used to calculate the total energy surface of the $^{208}$Pb nucleus in a multidimensional space, $E(R_1,a_1,w_1,R_2,a_2,w_2,\delta_2 )$. The seven variables of the total energy surface in the procedure used in this study are based on the d3pF formula. The differential evolution method, proposed by Storn and Price \cite{rev3}, is used to minimize the total energy in multidimensional space and then obtain the ground state binding energy and neutron density distribution parameters. The differential evolution method provides an efficient adaptive scheme for global optimization over continuous spaces and has been successfully used with the semi-microscopic approach to study the nuclear structure of $^{230}$Th \cite{rev4} and $^{288-308}$Og isotopes \cite{ref38}. All the degrees of freedom of the total energy are set to vary but some variables must be switched off to obtain the different density distributions. In the case of 2pF density distribution, only first two parameters are free and the rest of the parameters are set to zero, i.e. the minimization procedure consider $E(R,a,0,0,0,0,0)$ total energy surface. In the case of 3pF density distribution, only first three parameters are free and the rest of the parameters are set to zero, i.e. the minimization procedure consider $E(R,a,w,0,0,0,0)$ total energy surface. In the case of d3pF density distribution, all the seven variables are free.

There are currently dozens of parameterizations of the Skyrme force, and it may affect the inferred binding energy and/or neutron density distribution. The effect of Skyrme force has been evaluated in previous nuclear structure studies and the results indicated that the force used has a slight influence on the value of the calculated binding energy, while the inferred nuclear deformation is not affected \cite{ref40,rev4}. In order to select one Skyrme force for the conduct of the current study, previous studies were reviewed and SKM* was selected because its results match well with experimental values. Terasaki and Engel \cite{rev5}, for example, found that SKM* works better than SLY4 in the prediction of single-particle vibrational states calculated using the self-consistent method with quasiparticle random-phase approximation. In another study, Ismail et.al.\cite{ref40} found that the binding energies calculated by the semi-microscopic approach with SKM* force show small root mean square deviation from the experimental values compared to the results of SLy4 and SkP. The second reason for choosing SKM* force in this study is the presence of two previous studies that presented the nuclear density of $^{208}$Pb using the semi-microscopic approach with SKM* force but without restricting the proton density distribution \cite{ref12,ref40}. This gives an opportunity to compare and evaluate the efficiency of the improvements proposed in the current study without looking at the effect of the Skyrme force considered in the calculation.

In the case of 2pF distribution, the parameters considered in minimization are the radius parameter $R$ and the diffuseness parameter $a$. The minimum total energy is obtained at $R= 6.85$ fm and $a=0.85$ fm and has the value of $-1632.28$ MeV. In the case of 3pF distribution, the minimization is performed with respect to $R$, $a$, and $w$. The minimum total energy is obtained at $R=7.1$ fm , $a=0.48$ fm and $w=-0.27$ and has the value of $-1635.53$ MeV. As a result of increasing the degrees of freedom of minimization from 2 to 3, the nucleus gains extra binding energy of $3.25$ MeV. The neutron density distribution of $^{208}$Pb is not flat at the vicinity of the nucleus center, but it is raised. That is why considering central depression parameter $w$ adds to the binding energy of $^{208}$Pb. For other nuclei which are flat about the center, the effect of $w$ would be less significant. For light nuclei and ultra-heavy nuclei, the significance of central depression parameter would be high as the ground states of those nuclei are centrally raised and centrally depressed, respectively \cite{ref12}. In the case of d3pF, the minimum total energy is $-1636.66$ MeV and obtained at $R_1=6.77$ fm , $a_1=0.56$ fm, $w_1=-0.02$, $R_2=2.12$ fm , $a_2=0.47$ fm, $w_1=-2.49$ and $\delta_2=-0.047$. Increasing the degrees of freedom of minimization from 3 to 7 adds only $1.09$ MeV to the binding energy of $^{208}$Pb, about $1/3$ the gain in binding energy due to increasing the degrees of freedom from 2 to 3. Although the improvement in binding energy due to considering 7 parameters is relatively small, the shape of the density is improved strongly. D3pF distribution allows not only for central depression, but also for fluctuations in the nuclear interior. Moreover, comparison between the results of the three distribution formulae indicates that considering more parameters would not provide significant enhancement to the calculated binding energy.

The values of the parameters corresponding to the minimum binding energy of $^{208}$Pb are presented in Table \ref{tab1}, in addition to the values of rms radius of neutrons distribution, neutron skin thickness and the total energy. Comparing the calculated total energy values, corresponding to 2pF, 3pF and d3pF distributions, with the experimental total energy, we find that the difference between the calculated energy and the experimental value decreases with increasing the number of parameters. The smallest difference is obtained in the case of d3pF of 0.19 MeV. For 2pF and 3pF distributions, the calculated binding energy is smaller than the experimental binding energy by 4.15 MeV and 0.9 MeV, respectively. For d3pF, the calculated binding energy is greater than the experimental binding energy by 0.19 MeV. The experimental results of neutron distribution probing experiments include the values of rms radius, neutron skin thickness, or an interpolated fit to an analytical formula. The 2pF distribution parameters extracted from coherent pion photoproduction experiment, shown in Table \ref{tab1}, may give a good description of the tail of the density but not the detailed distribution near the center. Comparing the values of radius parameters of the three distributions considered in this study with the experimental value, it can be observed that the d3pF distribution gives the closest value to the experimental value, followed by 2pF and then 3pF. Likewise, the values of diffuseness parameters for both the d3pF and 2pF distribution agree with the experimental value, while the 2pF distribution shows less diffuseness. For this comparison, we consider the $R_1$ and $a_1$ parameters for the d3pF formula that describe the tail of the distribution.

\end{multicols}

\begin{center}
\tabcaption{ \label{tab1} The values of the minimum total energy and the corresponding density distribution parameters, in addition to neutron distribution rms radii and neutron skin thickness, for the 2pF, 3pF and d3pF formulae. The experimental values are mentioned for comparison.}
\footnotesize
\begin{tabular*}{12cm}{@{\extracolsep{\fill}}p{30 mm}|c|c|cc|c}
\toprule
		 		&	2pF		&	3pF		&\multicolumn{2}{c|}{d3pF}&	Exp.	\\
				& 			&			&	1		&	2		&			\\
\hline
$R$ (fm)		&	6.85	&	7.10	&	6.77	&	2.12	&  6.70 $\pm$ 0.03 \cite{ref24}\\
$a$ (fm)		&	0.52	&	0.48	&	0.56	&	0.47	& 0.55 $\pm$ 0.03 \cite{ref24}\\
$w$				&			&	-0.27	&	-0.02	&	-2.49	&			\\
$\delta$		& 			&			&	1.047	&	-0.047	&			\\
\hline
rms $R$ (fm)	& 	5.647	&	5.616	&	\multicolumn{2}{c|}{5.623}&$5.78_{-0.18}^{+0.16}$ \hphantom{00}\cite{rev1}\\
				& 			&			&			&		&$5.653_{-0.029}^{+0.026}$ \cite{rev2}\\
\hline
Neutron skin thickness 		&	0.144	&	0.113	&	\multicolumn{2}{c|}{0.120}&$0.15_{-0.06}^{+0.04}$ \hphantom{00}\cite{ref24}\\
		(fm)		& 			&			&			&		&$0.33_{-0.18}^{+0.16}$ \hphantom{00}\cite{rev1}\\
				& 			&			&			&		&$0.211_{-0.063}^{+0.045}$ \cite{rev2}\\
\hline
Total energy (MeV)&-1632.28	&-1635.53	&\multicolumn{2}{c|}{-1636.62}&-1636.43	\hphantom{000}\cite{rev6}\\
\hline
Diference from Exp. (MeV)&-4.15 &	-0.9&\multicolumn{2}{c|}{+0.19}&\\
\bottomrule
\end{tabular*}%
\end{center}

\begin{center}
\includegraphics[width=16cm]{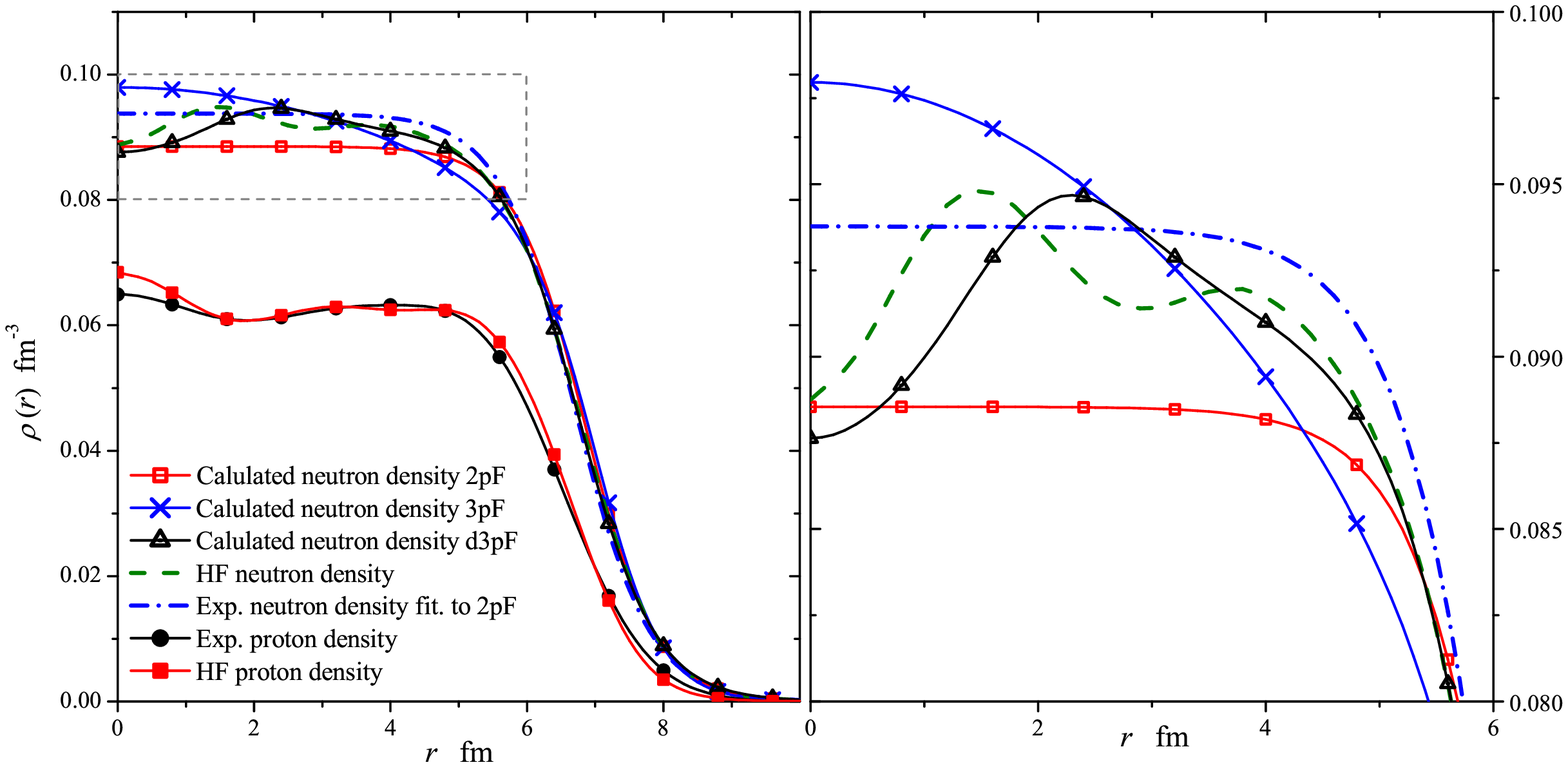}
\figcaption{\label{fig1}  The experimental proton density \cite{ref4} and calculated neutron density distributions of $^{208}$Pb. The proton and neutron density distributions based on the self-consistent HF calculation \cite{ref45}, and 2pF distribution based on pion photoproduction experiment \cite{ref24}, are also shown for comparison.}
\end{center}

\begin{multicols}{2}

The experimental data of rms radius and neutron skin thickness show relatively large uncertainty. Experimental data tell that the rms radius of $^{208}$Pb is between 5.6 fm and 5.94 fm \cite{rev1,rev2} and the neutron skin thickness is between 0.09 fm and 0.49 fm \cite{ref24,rev1,rev2}. The calculated values of rms radius and neutron skin thickness from the three resulting distributions are within these ranges, but generally give lower values than the mean values of the experimental results. More specifically, the resulting rms radius and neutron skin thickness considering 2pF neutron distribution are greater than the corresponding values in the case of d3pF and followed by 3pF. The experimental values, shown in Table \ref{tab1}, are extracted from pion photoproduction \cite{ref24}, parity violation experiment conducted at the Jefferson lab (PREX) \cite{rev1}, and proton elastic scattering \cite{rev2}. Experiments with electromagnetic probes have more clear theoretical basis than the strongly interacting probes and have the advantages of probing the full nuclear volume and less disturbing the ground state of the target nucleus. The experimental results are arranged in Table \ref{tab1} in ascending order according to the mass of the probing particle. For pion photoproduction experiment the incident particles are the photons, while in PREX parity violation experiment the electrons are used. The values of the neutron skin thickness calculated in the current study show compatibility with the data extracted from the pion photoproduction experiment, although the calculated values are less than the mean value, but all fall within the error range of the experimental result. The neutron skin thickness extracted from parity-violation and proton elastic scattering experiments are very large relative to that deduced by pion photoproduction and also the values calculated in this study. On the other hand, the values of the rms radius calculated in this study agree with the available practical values derived from the parity violation and proton elastic scattering experiments.

Fig.\ref{fig1} shows a comparison of calculated neutron densities obtained from total energy minimization with the neutron density obtained from the self-consistent Hartree-Fock (HF) calculation \cite{ref45} for the $^{208}$Pb nucleus. The choice of HF density for this comparison is based on the success of HF calculation to reproduce the experimental proton density, as shown in Fig.\ref{fig1}. Moreover, there is no available model independent experimental data for neutron density distribution of the $^{208}$Pb nucleus. The 2pF density is flat and almost constant around the center, while the 3pF density is raised at the center and decreases continuously and smoothly for the whole profile. The maximum density for both 2pF and 3pF density distributions is obtained exactly at the center of the nucleus. D3pF shows fluctuations around the center with maximum density at $r \approx 2.3$ fm. The three density distributions have smooth tails with almost the same values for radii larger than $6$ fm. The HF density has fluctuations around the center and shows maximum density at $r \approx 1.4$ fm.

For more precise comparison, differences were calculated between the calculated neutron densities and the corresponding HF density. Fig.\ref{fig2} shows the distribution of the differences for the 2pF, 3pF and d3pF formulae against the distance from the nuclear center. At the center, the 2pF distribution shows the smallest absolute difference, while the 3pF distribution shows the largest absolute difference. The range of difference, in fm$^{-3}$, between the 2pF and HF densities is $-0.0063:0.0028$ and between the 3pF and HF densities is $-0.0036:0.0092$. In the case of d3pF, the difference range is almost symmetric about zero and has the smallest length. The distribution of differences can be better understood by statistical analysis, because it reflects abstract information from all data points, not just from characteristic points. Table\ref{tab2} shows some statistical parameters for the differences between calculated densities and HF densities based on the data generated for r values up to $12.0$ fm in steps of $0.1$ fm. The maximum and minimum differences give the extreme values of difference for each dataset, which can be clearly seen from Fig.\ref{fig1}.

\begin{center}
\includegraphics[width=8.5cm]{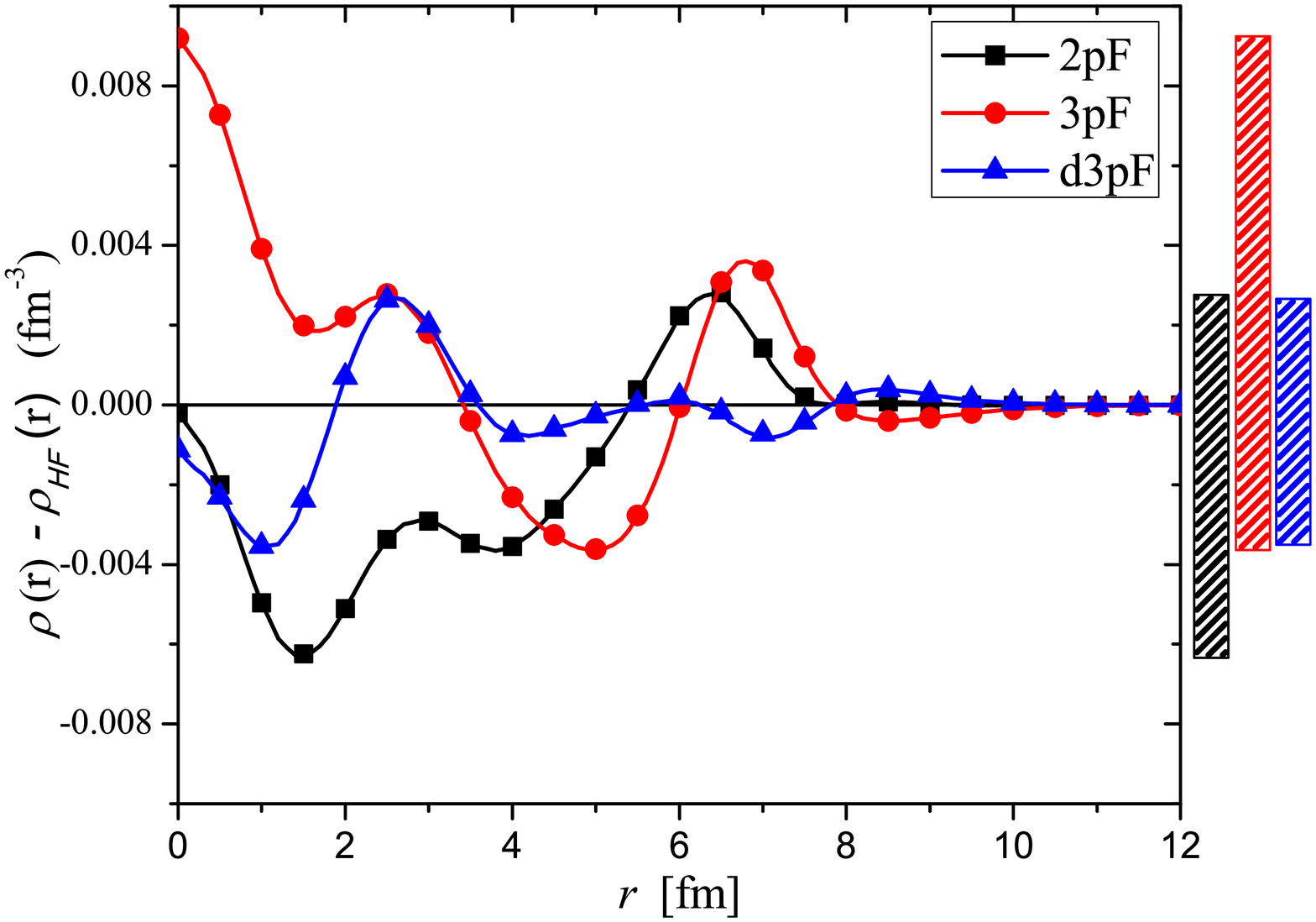}
\figcaption{\label{fig2}  The differences between calculated neutron density distributions and the HF neutron density distribution. The three bars on the right represent the difference between the highest and lowest values of differences corresponding to 2pF, 3pF and d3pF distributions, respectively from left to right.}
\end{center}

However, extreme values cannot indicate the average difference or total difference between the HF distribution and any of the distributions. The mean difference for the 2pF density is $-1.20 \times 10^{-3}$  fm$^{-3}$, which indicates that it predicts lower density than the HF density for the most of data points. For the 3pF, the mean difference is $7.82 \times 10^{-3}$  fm$^{-3}$, which indicates that it predicts higher density than the HF density for the most of data points. The magnitude of mean difference for the 3pF distribution is more than 6 times larger than for 2pF, which indicates that the density predicted by the 3pF formula is, in average, more far from the HF density than the density predicted by the 2pF formula. The mean difference for the d3pF density is $-2.25 \times 10^{-4}$  fm$^{-3}$, which is closer to zero than the other two densities. This means that the d3pF formula predicts a neutron density close to the HF density, in terms of values and shape, better than 2pF and 3pF. The sum of the differences for each dataset shows how the data points are distributed on both sides of the zero difference. The closer the sum of the differences to zero the more symmetric distribution about the zero. The values of sum of the differences indicate that the differences in the case of d3pF distribution are more symmetric about zero than the other two distributions.

\begin{center}
\tabcaption{ \label{tab2} The minimum, maximum, mean, sum and sum of squares of differences between the neutron densities calculated by 2pF, 3pF and d3pF formulae and the corresponding HF density.}
\footnotesize
\begin{tabular*}{85 mm}{@{\extracolsep{\fill}}p{5 mm}ccccp{10 mm}}
\toprule &Minimum 	&	Maximum 	&	Mean 	&	 sum	&	 sum of squares	\\
\hline
2pF	&-6.27E-3&	2.81E-3&	- 1.20E-3			&	-0.14518			&	8.09E-4\\
3pF &-3.63E-3&	9.20E-3&	\hphantom{0}7.82E-4&	\hphantom{0}0.09457&	9.77E-4\\
d3pF&-3.54E-3&	2.67E-3&	-2.25E-4			&	-0.02726			&	1.90E-4\\
\bottomrule
\end{tabular*}%
\end{center}

The sum of the differences cannot indicate whether the difference values are, in total, large or small. In order to estimate the total difference, the amount of dispersion about the zero difference can be assessed by calculating the differences sum of squares. The differences sum of squares reflects the total absolute difference regardless of the direction. The values of differences sum of squares for 2pF, 3pF and d3pF are $8.09 \times 10^{-4}$, $9.77 \times 10^{-4}$ and $1.90 \times 10^{-4}$, respectively. This implies that the d3pF density shows the lowest dispersion about the HF density, followed by 2pF and then 3pF. Although the 3pF formula enhances the calculation of total energy compared to 2pF, the 2pF formula shows a better distribution for the $^{208}$Pb nucleus compared to the HF density. The d3pF formula outperformed the other two formulae in estimating the total energy and inferring a better density distribution.

\section{Summary and Conclusion}
In the present study, the neutron density distribution of the $^{208}$Pb nucleus is investigated assuming that the proton density distribution is constrained to its experimental distribution. The total energy surface in a seven dimensional space is calculated in the framework of semi-microscopic approach based on the Skyrme interaction and Strutinsky's shell and pairing corrections. The neutron density distribution is considered in the form of 2pF, 3pF and d3pF parameterized distributions. The ground state binding energy and the neutron density distribution parameters are obtained from the energy surface for the three distributions considered in the study. As the number of degrees of freedom in the density formula increases, the calculated total energy decreases. The difference between the total energies calculated considering 2pF and 3pF formulae is $3.25$ MeV, while the difference between the total energies calculated considering 3pF and d3pF formulae is $1.09$ MeV. Although the added energy due to considering d3pF formula in relatively small, it is very important because it allows for central fluctuation.

Comparing the three results with the experimental total energy, we find that the difference between the calculated energy and the experimental value decreases with increasing the number of parameters. The smallest difference is obtained in the case of d3pF of $0.19$ MeV. The calculated values of rms radius agree with the experimental values of the parity-violation and proton scattering experiments, while the neutron skin thickness values were smaller than the experimental values. On the other hand, the inferred d3pF neutron density distribution agrees with the pion photoproduction experiment in the neutron skin thickness and the diffused part of the distribution. Comprehensive comparison of the calculated neutron densities with  HF density showed that replacing 2pF with 3pF does not improve the overall density distribution predicted for the $^{208}$Pb nucleus, however it improves the calculated total energy. Considering d3pF density improves both the calculated total energy and the neutron density distribution.

The main objective of this study is to present the idea of using reliable experimental findings to guide the results of theoretical calculations. The same principle can be applied to other theoretical models from the simplest to the most complex. Applying this principle on the semi-microscopic approach in the present study fairly improved the neutron density of $^{208}$Pb neucleus and the ground state total energy. The approach of constraining to experimental values may yield more promising results with other models, but this requires extensive studies to investigate its significance.

\end{multicols}

\vspace{-5mm}
\centerline{\rule{80mm}{0.1pt}}
\vspace{0mm}

\begin{multicols}{2}

\end{multicols}

\clearpage

\end{CJK*}
\end{document}